# Compression Behavior of Single-layer Graphene

*Otakar Frank[1,†], Georgia Tsoukleri[1,2], John Parthenios[1,2], Konstantinos Papagelis[3], Ibtsam Riaz[4], Rashid Jalil[4], Kostya S. Novoselov[4,\*], and Costas Galiotis[1,2,3,♣]*

[1]Institute of Chemical Engineering and High Temperature Chemical Processes, Foundation of Research and Technology-Hellas (FORTH/ICE-HT), Patras, Greece

[2]Interdepartmental Programme in Polymer Science and Technology, University of Patras, Patras, Greece

[3]Materials Science Department, University of Patras, Patras, Greece

[4]School of Physics and Astronomy, University of Manchester, Manchester, UK

[†] on leave from J. Heyrovsky Institute of Physical Chemistry, v.v.i., Academy of Sciences of the Czech Republic, Prague 8, Czech Republic

[\*] kostya@manchester.ac.uk

[♣] c.galiotis@iceht.forth.gr

ABSTRACT

Central to most applications involving monolayer graphene is its mechanical response under various stress states. To date most of the work reported is of theoretical nature and refers to tension and compression loading of model graphene. Most of the experimental work is indeed limited to bending of single flakes in air and the stretching of flakes up to typically ~1% using plastic substrates. Recently we have shown that by employing a cantilever beam we can subject single graphene into various degrees of axial compression. Here we extend this work much further by measuring in detail both stress uptake and compression buckling strain in single flakes of different geometries.  In all cases the mechanical response is monitored by simultaneous Raman measurements through the shift of either the G or 2D phonons of graphene. In spite of the infinitely small thickness of the monolayers, the results show that graphene embedded in plastic beams exhibit remarkable compression buckling strains. For large length (l)-to-width (w) ratios ($\geq$ 0.2) the buckling strain is of the order of -0.5% to -0.6%. However, for l/w <0.2 no failure is observed for strains even higher than -1%.  Calculations based on  classical Euler analysis show that the buckling strain enhancement provided by the polymer  lateral support is more than six orders of magnitude compared to suspended graphene in air.

Graphene is a two-dimensional crystal, consisting of hexagonally-arranged covalently bonded carbon atoms and is the template for other carbon allotropes.[1, 2] Graphene exhibits a high level of stiffness and strength with Young's modulus values of about 1TPa and strength in excess of 160GPa.[3, 4] It also possesses unique electronic properties, which can be further effectively modified by stress/strain.[5, 6] In fact, strain engineering has been proposed as a route for developing graphene circuits[7] and, in this respect, a precise determination and monitoring of stress and strain are key requirements. Furthermore, there is a growing interest in the exploitation of graphene as a nano-reinforcement in polymer based composites[8-10] for which it is important to know how efficiently the external stress is transferred from the matrix to the nano-inclusions.

Probing the shift of phonon frequencies is an effective way of assessing the degree of stress transfer of a material under an applied stress or strain along a given axis. Raman spectroscopy has proven very successful in monitoring phonons of a whole range of graphitic materials including graphene under uniaxial stress[11-16] or hydrostatic pressure.[17, 18] We have recently shown that the position of the 2D peak, $\omega_{2D}$, is related to the applied uniaxial strain, $\varepsilon$, at a rate of approximately $-65 \times 10^{-2}$ cm$^{-1}$.[13, 16] Past reports of much lower shifts by a number of authors have been attributed[16] to the effect of substrate and/or to the presence of residual strain in the monolayer. The dependence of the G peak position under uniaxial strain has also been the subject of intense interest and, as in the case of the 2D peak, substantial discrepancies have been reported in the literature.[12-14] In the recent work reported by us [13] significant G peak splitting is observed due to the lowering of the $E_{2g}$ phonon symmetry by the imposition of a uniaxial strain.

With a few notable exceptions (see above and e.g. [19-22]), most works dealing with mechanical properties of graphene (see e.g. [6, 7, 23-25]) are of theoretical nature and generally limited to suspended graphene at the atomic scale. Hence, there is a growing demand for experimental data to validate the models and relate them to graphene attached to various substrates. In the present work, graphene flakes are subjected to a cyclic uniaxial deformation (tension - compression) using the polymer cantilever beam technique. The effect of compressive strain on the doubly degenerate G Raman band is presented for the first time. It was found that for compressive strain of about -0.1% the G band is split in a fashion similar to that observed in tension.[13] The critical strain for graphene buckling was found to be dependent on the flake size and geometry with respect to the strain axis and as such it follows the classical Euler buckling behavior. However, the role of substrate is found to be of a crucial importance, by enhancing the critical buckling strain by several orders of magnitude compared to suspended flakes. Finally, by employing the strain sensitivity of the 2D Raman band *post mortem* strain maps of the flake were constructed. The strain topography on these maps reveals a wrinkling pattern which is established on the flake on the completion of the cyclic deformation. Such patterns are found to be dependent on both the strain axis direction and the flake aspect ratio; a result that should be taken into account in applications such as all-graphene circuits.[6]

RESULTS AND DISCUSSION

Graphene monolayers were subjected to compressive and tensile loading by means of a cantilever beam assembly (Fig. 1a). The specimens were embedded into two polymeric layers of SU8 and S1805 and placed onto PMMA bars (Fig. 1a and SI). A detailed description of the experimental set-up and the sample preparation procedure are presented in SI and ref. 16. Raman sampling was performed *in situ* on different sample locations depicted with crosses in Figs. 1b-d.

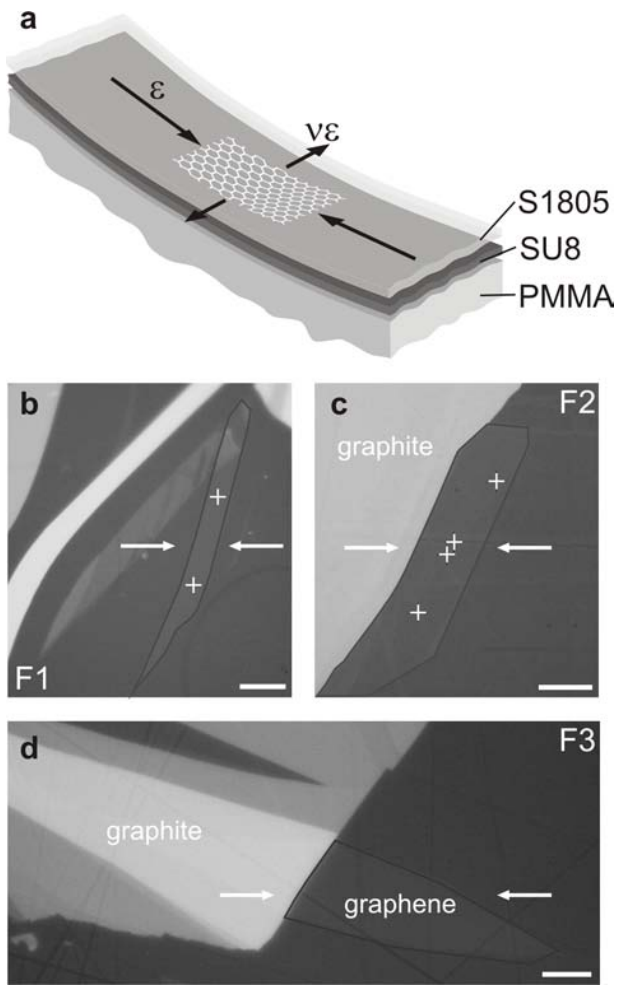

**Figure 1.** A scheme of the beam bearing the graphene sample under study (a). Optical micrographs of the graphene flakes investigated; flake F1 (b), flake F2 (c) and flake F3 (d). The scale bar is 10 μm and the arrows indicate the strain axis. The crosses in (b) and (c) represent sampling locations.

Figure 2 shows representative Raman spectra of a graphene monolayer in the G peak region as a function of strain recorded on the flake F1 (shown in Fig. 1). Positive (negative) strain values denote data taken under tension (compression). As seen in Fig. 2, the doubly degenerate $E_{2g}$ optical mode (G peak) splits into two components, which have been termed[12, 13] $G^-$ and $G^+$ in analogy with nanotubes, referring to polarization along the strain and perpendicular to it, respectively.[12, 13] The most striking feature in Fig.2 is the G peak splitting under both tension and compression; in both cases the $E^+_{2g}$ phonon is perpendicular to the applied strain and thus experiencing smaller softening (redshift) or hardening (blueshift) whereas the $E^-_{2g}$ being parallel to strain is showing much greater rates of shifting in all cases. The rate of shifting of both modes is affected by the Poisson's ratio $\nu=0.33$ [13] of the substrate, assuming ideal adhesion between the flake and the polymer matrix. The $G^-$:$G^+$ intensity ratios remain relatively constant during the course of loading and are the same for all investigated spots on a particular flake, being 1.5:1 for F2 flake and 1:1 for F1 flake. The difference between the two flakes is caused by their different crystallographic orientation with respect to the strain axes.[13] The $G^-$ and $G^+$ polarization angle dependence is described in detail in refs 12 and 13.

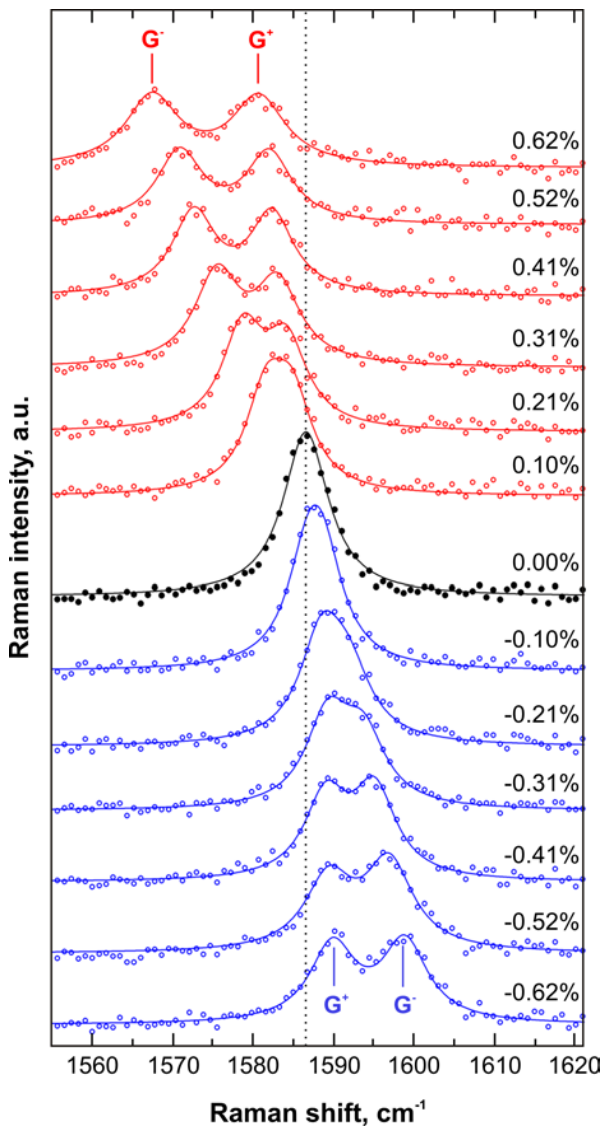

**Figure 2.** G band Raman spectra of graphene flake excited at 785 nm under uniaxial strain (positive values for tensile and negative for compressive strain). Data were recorded around the center of the flake F1. The original measurements are plotted as points. The solid curves are the best Lorentzian fits to the experimental spectra.

In Figure 3a the G$^-$ and G$^+$ peak positions (further denoted as Pos(G$^-$, G$^+$)) as a function of the compressive strain are shown for flakes F1 and F2. The Pos(G) at zero strain and the slopes $\partial\omega_G^+/\partial\varepsilon$ and $\partial\omega_G^-/\partial\varepsilon$ for all specimens and different experiments are summarized in Table S1 (Supporting Information). The sensitivity of the individual G bands is higher under tension (Table 1), being -31.4 ± 2.8 cm$^{-1}$/% for the G$^-$ mode and -9.6 ± 1.4 cm$^{-1}$/% for the G$^+$. Under compression, the average sensitivities for the two specimens differ. The F1 flake shows 5.5 ± 1.9 cm$^{-1}$/% for the G$^+$ mode and 22.3 ± 1.2 cm$^{-1}$/% for the G$^-$ mode, while the F2 flake exhibits 10.1 ± 2.1 and 33.1 ± 2.2 cm$^{-1}$/% for G$^+$ and G$^-$ modes, respectively. The flake F2 shows $\partial\omega_G/\partial\varepsilon$ values in the linear part of the curves close to zero strain similar to tension, while the F1 flake sensitivities are by ~ 30 % lower. The values extracted in the present study under both tension and compression are given in Table 1. For comparative purposes the reported values in the literature for the slopes $\partial\omega_G/\partial\varepsilon$ under tension are also included.

The issue of residual strain present in the embedded flake is of paramount importance for the mechanical behaviour of graphene as has been shown previously.[16, 26, 27] Especially for the embedded graphene into polymer matrices, the residual strain is due to either the initial deposition process and/or the shrinkage of resin during solidification (curing). The roughness of the polymer substrate may also play a role. The laser Raman technique employed here allows us to identify the presence of residual strain by just measuring the Raman frequency of the embedded flake and compare it to that of an unstressed flake or literature value (e.g. 2680 cm$^{-1}$ for laser excitation at 514 nm). In this work, in order to eliminate the effect of residual strain upon the mechanical data, we selected flakes that exhibited zero or minimal residual strains following a two step methodology. In the first step a Raman mapping is performed that covers a broad area of the flake. The 2D Raman band is then used to generate two separate contour maps whereas the first one presents the topography of the Pos(2D) on the flake and the other the full-width-at-half-maximum (FWHM) of the same flake locations. Based on the fact that the FWHM of the 2D Raman band increases with deformation, the minimum residual strain regions can then be identified by correlating the two topographies; these are the regions where the topography exhibits minimum FWHM values. Even though it is practically impossible to obtain an absolutely prestrain-free monolayer, the small variations in the initial band frequencies observed in our experiments do not seem to affect the measured $\partial\omega/\partial\varepsilon$ at the particular spots. Furthermore, the low prestrain level is evidenced by the linear response of the band sensitivities to tension. As shown previously,[16] a pre-compression would be accompanied by a lower starting $\partial\omega/\partial\varepsilon$ value and a parabolic $\omega(\varepsilon)$ dependence.

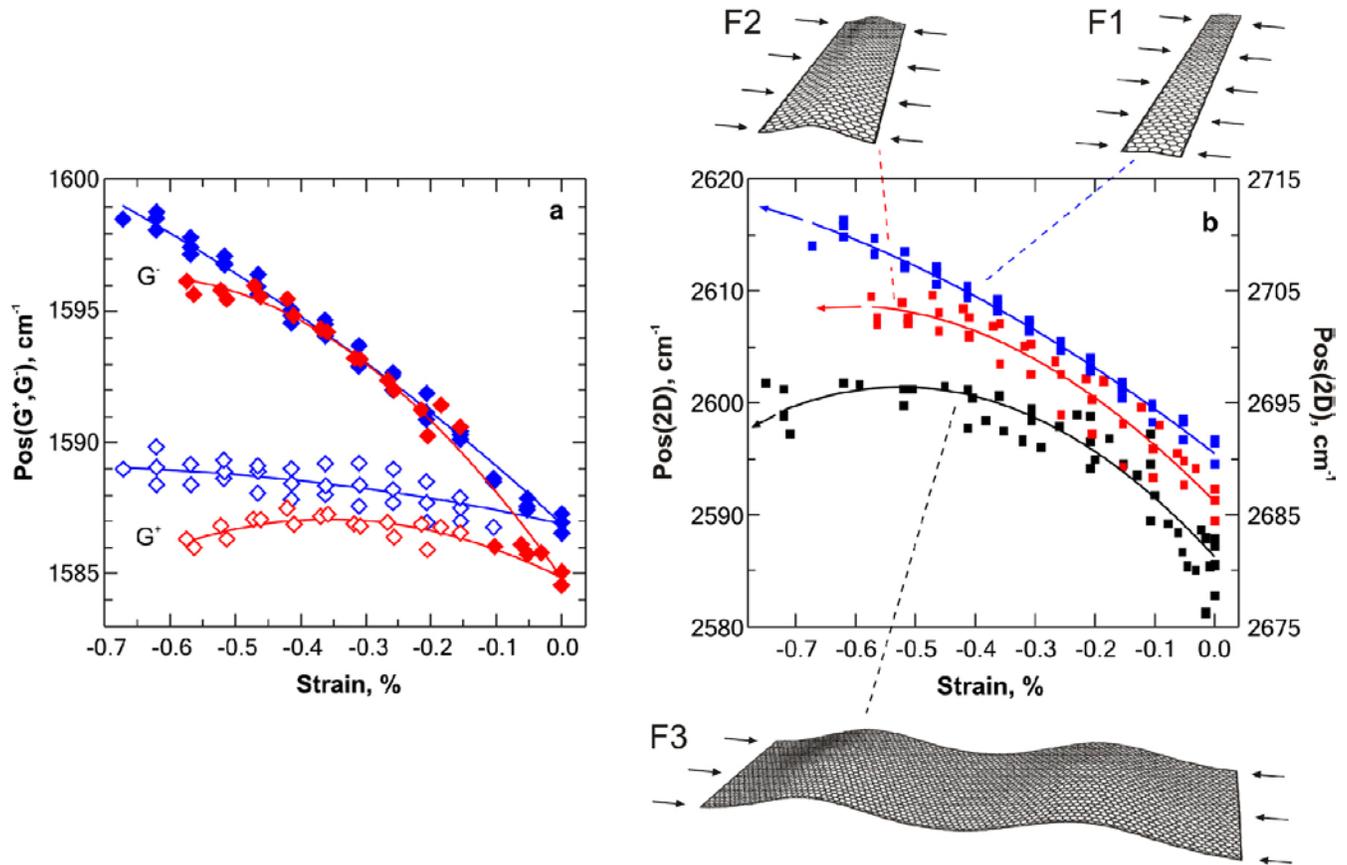

**Figure 3.** (a) The splitting of G band under compressive strain for F1 (blue) and F2 (red) graphene flakes. Empty and full diamonds indicate the frequency of the $G^+$ and $G^-$ sub-bands, respectively. Solid lines represent 2$^{nd}$ order polynomial fits where all measurements on a specific flake has been taken into account. (b) Pos(2D) as a function of compressive strain for graphene flakes with different orientations. Blue and red squares belong to F1 and F2 flake, respectively, and are plotted against the left axis. Black squares indicate Pos(2D) for F3 flake and are plotted against the right axis. Data for F3 flake are acquired using 514 nm excitation and reproduced from ref. 16. Solid lines represent second order polynomial fits to the experimental data. The corresponding graphene flakes are schematically illustrated as rectangular shells with aspects ratios ($l/w$) that correspond to the real ones and schematically indicate the number of half-waves generated by compression (see text). Arrows indicate the compression axis.

**Table 1.** Summary of $\partial\omega_G/\partial\varepsilon$ values in tension and compression.

| | Compression | | Tension | |
|---|---|---|---|---|
| | $\partial\omega_G^-/\partial\varepsilon$ | $\partial\omega_G^+/\partial\varepsilon$ | $\partial\omega_G^-/\partial\varepsilon$ | $\partial\omega_G^+/\partial\varepsilon$ |
| | cm$^{-1}$/% | cm$^{-1}$/% | cm$^{-1}$/% | cm$^{-1}$/% |
| [14] | --- | --- | -14.2 | |
| [12] | --- | --- | -12.5 ± 2.6 | -5.6 ± 1.2 |
| [13] | --- | --- | -31.7 | -10.8 |
| This work | 22.3 ± 1.2 [a, *] | 5.5 ± 1.9 [a, *] | -31.4 ± 2.8 | -9.6 ± 1.4 |
| | 33.1 ± 2.2 [b, *] | 10.1 ± 2.1 [b, *] | | |

\* the values correspond to the linear part close to zero strain level of the $\omega_G(\varepsilon)$ curves

[a] – flake F1, [b] – flake F2

In tension, the Raman wavenumbers of the $E^-_{2g}$ and $E^+_{2g}$ sub-bands follow almost perfectly linear trends up to the maximum applied strain. However, in compression the linearity holds for strain levels up to 0.3-0.5%. As shown in Fig. 3a, Pos(G$^+$) of F2 reaches a plateau at a strain value of 0.4%, while the $\partial\omega_G^+/\partial\varepsilon$ of F1 remains almost constant. Similar differences in behavior of the two flakes can be also detected in the corresponding $\omega_G^-(\varepsilon)$ curves. It is worth noting here that the slopes $\partial\omega_{G}^{-,+}/\partial\varepsilon$ in compression evaluated for different mapping locations on a particular flake show small differences that can be attributed to inhomogeneities of the strain field within the flake (Table S1, Supporting Information).

A further insight into the compressive behavior of graphene is provided by the Pos(2D) dependence on compressive strain by comparing previously reported data[16] acquired using an excitation laser line at 514 nm. In Fig. 3b, three distinct sets of experiments for each particular graphene flake are presented. Similarly to the compressive behavior of the G band, Pos(2D) exhibits a non-linear trend with strain for all flakes which can be captured by second order polynomials. The observed $\partial\omega_{2D}/\partial\varepsilon$ is ~+55 cm$^{-1}$/ % and ~+42 cm$^{-1}$/ % for flake F2 and F1, respectively, at zero strain. For comparison it is recalled that the $\partial \omega_{2D}/\partial \varepsilon$ measured previously using an excitation laser line at 514 nm was ~+59 cm$^{-1}$/%.[16] Interestingly it should be noted that in all flakes Pos(2D) relaxes after an abrupt uptake. The onset strain of the Pos(2D) relaxation is at different value for each flake.

The moment of the final failure of the flakes can be expressed by the critical buckling strain ($\varepsilon_c$). For comparison purposes between flakes, we define $\varepsilon_c$ as the local maxima in the 2$^{nd}$ order polynomials fitted to Pos(2D) vs. strain values. The $\varepsilon_c$ value for F1 flake can be only extrapolated from the polynomial, giving 1.25%. For F2 and F3 flakes which showed clear failure, the $\varepsilon_c$ values were estimated at 0.53% and 0.64%, respectively. All compression data are summarized in Table 2.

The critical buckling strain for a flake in the classical Euler regime in air, can be determined through the following equation:[28]

$$\varepsilon_c = \frac{\pi^2 kD}{Cw^2} \quad (1)$$

where $w$ is the width of the flake, $k$ is a geometric term (see below), and D and C are the flexural and tension rigidities, respectively. A tension rigidity value of 340 GPa nm has been reported by AFM[3] measurements whereas the flexural rigidity has been estimated to 3.18 GPa nm$^3$.[5, 16] The above equation (1) is mainly valid for suspended thin films and yields extremely small (~10$^{-9}$) $\varepsilon_c$ values for graphene monolayers of thicknesses of the order of atomic radii. Such extremely small critical buckling strains are also predicted by molecular dynamics calculations.[24] However for embedded flakes the above predictions are meaningless since current and previous experimental results[16] clearly point to much higher values of strain prior to flake collapse.

When embedded in a polymer matrix, the graphene is prevented from full buckling due to the lateral support offered by the surrounding material. At a certain strain the interface between graphene and polymer should weaken or fail and the flake may buckle as it would do in air. Therefore, assuming that $\varepsilon_c \propto \frac{k}{w^2}$, the different response of the individual graphene flakes to compression can be determined by their geometries and orientation with respect to the strain axis. The geometric term *k* is dependent on the aspect ratio combined with a number of half waves *m* into which the flake buckles:[28]

$$k = \left(\frac{mw}{l} + \frac{l}{mw}\right)^2 \quad (2)$$

For the F3 flake, where length ($l$) = 56 μm and width ($w$) = 25 μm, 3 half waves are expected to occur at the critical load,[28] thus $k_{F3}$ = 4. For flakes F2 and F1, where $l/w < 1$, only one half wave appears, thus $k_{F2}$ = 22.7 and $k_{F1}$ = 89.1. The number of half-waves is illustrated on the respective sketches in Fig. 3b. Accordingly the term $\frac{k}{w^2}$ increases from 0.006 μm$^{-2}$ for F3 flake up to 0.028 μm$^{-2}$ for F1 flake (Table 2). If we now plot the $\frac{k}{w^2}$ as a function of $\varepsilon_c$, a linear dependence for the three studied flakes is obtained (Figure 4). The equation of the least-squares-fitted line is given by:

$$\frac{k}{w^2} = a\varepsilon_c + b \quad (3)$$

where the slope $a$ = -0.03 μm$^{-2}$.

Since as shown in Figure 4, an Euler type analysis can be applied to the embedded graphene then the critical buckling strain should be given by:

$$\varepsilon_c^{embedded} = \frac{k}{w^2}\frac{D^*\pi^2}{C} \quad (4)$$

where $D^*$ is now the flexural rigidity in the presence of the polymer. With reference to the slope $a$ = -0.03 μm$^{-2}$ in Figure 4, the $D^*$ can be estimated to 12 MPa μm$^3$, which is, indeed, 6 orders of magnitude higher than the value in air. This is truly a remarkable finding that indicates clearly that the support offered by polymer barriers to a rigid monolayer can provide a dramatic enhancement to its compression behavior. The recently published results[10] showing measurable improvements in the compression behavior of polymers by the addition of graphene at low volume fractions also confirm our findings here. The effect of lateral support can also be deducted from our previously reported results[16] involving a graphene flake of dimensions, $l$ = 8 and $w$ = 6 μm, simply laid on top of a substrate. As was shown in ref. 16, the measured $\partial\omega_{2D}/\partial\varepsilon$ of 25 cm$^{-1}$/% at zero strain is 2-2.5 times lower than the value expected (this work, Figure 4) for a fully supported flake.

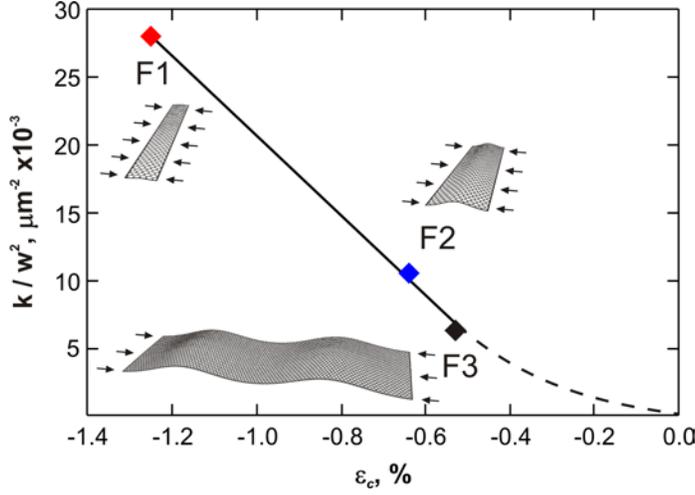

**Figure 4.** Geometrical term $\frac{k}{w^2}$ plotted against critical buckling strain $\varepsilon_c$ for the three flakes under study. The solid line represents a line fit to the obtained experimental points. The dashed line shows the possible evolution of $\frac{k}{w^2}(\varepsilon_c)$ when $\frac{k}{w^2}$ limits to zero.

**Table 2.** Critical buckling strain ($\varepsilon_c$), geometrical terms $k$ and $k/w^2$, and approximate physical dimensions (length $l$ and width $w$, with $l$ oriented along the strain axis) of the studied graphene flakes.

| Sample | $\varepsilon_c$ (%) | $k/w^2$ (μm$^{-2}$) | $k$ | $l$ (μm) | $w$ (μm) |
|---|---|---|---|---|---|
| F1 | -1.25 | 0.028 | 89.12 | 6 | 56 |
| F2 | -0.64 | 0.011 | 22.71 | 11 | 50 |
| F3 | -0.53 | 0.006 | 4.02 | 56 | 25 |

It has to be noted that the above described approach of defining the influence of the support, and hence the interaction between the substrate and the graphene flake, by a single term $D^*$ is very simplified. Ideally, different stages of the compression process need to be addressed separately, as described e.g. in [29], to quantify the effect of debonding first, followed by the buckling itself. However, the use of common phenomenological models is unsatisfactory given the unique nature of 2D membranes one atom thick, yet macroscopic in lateral dimensions.[30] Similarly, the Euler type buckling observed in the studied flakes is not necessarily universal in the whole $\frac{k}{w^2}$ range. As can be seen in Fig.4, the fitted line does not pass through zero which indicates that its validity for $\frac{k}{w^2} < 5 \times 10^{-3}$ μm$^{-2}$ is questionable. In the other extreme case where $l \gg w$ and $w$ is in the nanometer scale, a non-linear behaviour governed by the matrix effects can be expected too.[31] A further study of this $\frac{k}{w^2}$ region will be essential to assess the mechanical properties of graphene nanoribbons.

Now we come to the FWHM of the peaks studied which provides valuable complementary information on the structural changes in the flake that occur during mechanical loading. The Figure S3 (Supporting Information) shows the G band behavior under compression for a spot in the flake F2. A linear increase of Pos(G$^+$,G$^-$) with strain can be observed up to -0.35%, where a subsequent relaxation of the Raman shift values takes place. The FWHM, which is equal for both sub-bands at a given strain level in the whole strain range measured, follows a different evolution. At first, it increases at a rate lower than 2 cm$^{-1}$/% and, then, at a strain level of -0.5% increases rapidly, reaching values over 10 cm$^{-1}$ at -0.6%. The rate of broadening in the final stage exceeds 25 cm$^{-1}$/%. Exactly the same behavior, i.e. rapid broadening at the onset of failure was observed in all compression experiments on flake F2. In contrast, the F1 flake does not show a pronounced FWHM(G) increase. This is in accordance with the almost linear slope of the $\omega_G(\varepsilon)$ curves in F1 sample and the negligible increase of the FWHM(G) under tension, which is less than 2 cm$^{-1}$/%. Similar dramatic G band broadening is observed on buckled graphene suspended over a trench designed in silicon substrate.[32] In that case, a compression is induced by heating and subsequent shrinkage of graphene due to different thermal expansion coefficient compared to underlying silicon.[32]

Figure 5 shows recorded Raman maps from the central part of specimen F2 at rest on the completion of the cyclic loading. Strain levels in Fig. 5a were calculated using the 2$^{nd}$ order polynomial fitted to the Pos(2D) data of the flake F2 as shown in Fig. 3b. Both the 2D band position (Fig. 5a) and FWHM (Fig. 5b) are presented. From Fig. 5a it can be deduced that most of the flake area is under a compressive strain up to -0.3%. As can be clearly seen in Figs 5a and b, Pos(2D) and FWHM(2D) maps point to a graphene monolayer which is not perfectly flat or at least with an inhomogeneous strain distribution. Indeed, a careful examination of the maps reveals areas with either maximum or minimum Pos(2D) but a significant band broadening in all cases. This is a clear indication of permanent wrinkling formation in the *post-mortem* flakes. Regarding the wrinkling pattern, the orientation of the longer axes of the FWHM isolines (Fig.5b) is approximately parallel to the edge of the neighboring bulk graphite. In the Pos(2D) map, the orientation of the isolines is similar, though more perturbed on the right edge of the graphene flake. The angle between the strain axis and the graphite edge, of about 50°, affects the direction of the formed wrinkles during the loading experiments. The graphite, thus, can act as a "clamp" for controlling the orientation of the wrinkles, which could be a key factor for tailoring the strain field characteristics in graphene-based electronic devices.

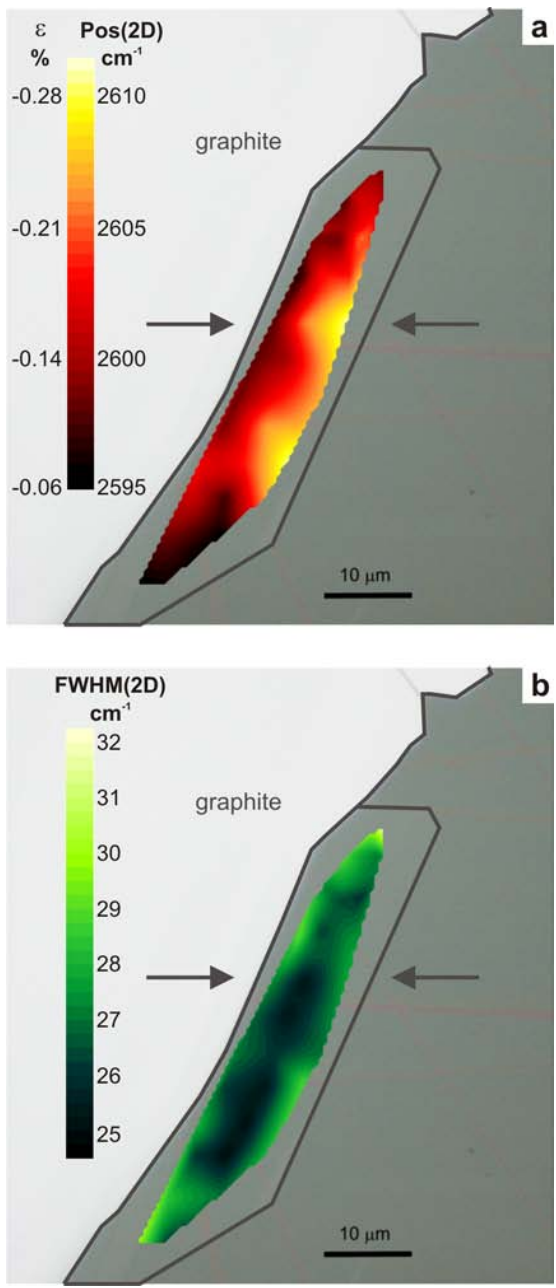

**Figure 5.** Post mortem (a) Pos(2D) and (b) FWHM(2D) maps of specimen F2 after cyclic loading. The light grey area in both (a) and (b) corresponds to bulk graphite. The arrows indicate the strain direction. See also Figure S4 (Supporting Information) for Pos(G) and FWHM(G) maps on the same flake.

CONCLUSIONS

In summary, we documented in detail the response of graphene monolayers to uniaxial strain by probing its optical phonons by Raman spectroscopy. In order to present a complete picture, frequency and FWHM of both G and 2D bands were monitored during tension and compression cycles. Flakes that exhibited minimum residual strain were selected by preliminary mapping. In addition, the linearity of the G and 2D bandshift with tensile strain further confirmed the low pre-strain level of selected flakes. However, in compression the G and 2D band response is non-linear and varies from flake to flake. The corresponding $\partial \omega_{G,2D}/\partial \varepsilon$ values decrease with strain till the eventual turn-over of the slope, which is indicative of progressive buckling that precedes the final collapse of the flake. The gradual decrease of $\partial \omega_{G,2D}/\partial \varepsilon$ is accompanied by an abrupt broadening of the bands, observed particularly in the G mode. The estimated critical buckling strain has been found to depend on size and geometry as would do any thin plate in an Euler buckling regime. It has to be stressed that the critical strain values of the embedded graphene flakes are remarkably high compared to the suspended ones. However, the effect of the lateral support provided by the polymer matrix is indeed dramatic and increases the effective flexural rigidity of graphene by 6 orders of magnitude. Finally, a post-mortem mapping of the flake indicates the presence of permanent wrinkles at an angle dictated by the neighbouring bulk graphite, which acts as a "clamp" supporting one edge of the compressed graphene.

METHODS

Graphene monolayers were prepared by mechanical cleavage from natural graphite (Nacional de Grafite) and transferred onto the PMMA cantilever beam covered by a ~200 nm thick layer of SU8 photoresist (SU8 2000.5, MicroChem). After placing the graphene samples, a thin layer of S1805 photoresist (Shipley) was spin-coated on the top. The beam has a total thickness of $t$ = 2.9 mm and width $b$ = 12.0 mm. The graphene flake was located at a distance, $x$, from the fixed end of 12.97 and 12.72 mm, resp.

The top surface of the beam can be subjected to a gradient of applied strain by flexing the beam by means of an adjustable screw positioned at a distance $L = 70.0$ mm from the fixed end. The deflection $\delta$ was measured accurately using a dial gauge micrometer attached to the top surface of the beam. The validity of this method for measuring strains within the -1.5% to +1.5% strain range has been verified earlier.[33]

MicroRaman (InVia Reflex, Rensihaw, UK) spectra are recorded with 785 nm (1.58eV) excitation, while the laser power was kept below 0.85 mW to avoid laser induced local heating on the sample. A 100x objective with numerical aperture of 0.9 is used, and the spot size is estimated to be ~1x2 μm. The polarization of the incident light was kept parallel to the applied strain axis. Because the graphene peaks overlap with strong peaks originated from the substrate, the spectra were first baseline (linear) subtracted, then normalized to its most intense peak of the substrate at 1450 cm$^{-1}$, and subsequently the spectrum of bare substrate was subtracted. Figure S1 (Supporting Information) shows the original spectra of bare substrate and unstressed graphene in the G band region, the same free graphene is then shown "as clean" in Fig. 2. All bands in the Raman spectra of graphene were fitted with Lorentzians. The FWHM of the G band for the unstressed graphene was found to be approximately 6-8 cm$^{-1}$.

The excitation wavelentgh (785 nm) was chosen with respect to a fluorescence of the polymer matrix embedding the graphene flakes. The fluorescence rendered measurements with lower excitations impossible or at least very difficult. In spite of a lower sensitivity of the CCD camera at higher wavenumbers, the 2D band is still clearly observable and can be evaluated, when the spectra are excited with 785nm laser line. The amplitudes of G and 2D bands of a graphene monolayer are approximately equal in this case. The FWHM of the 2D band in unstrained flakes was 24-25 cm$^{-1}$. The 2D linewidths and lineshapes, together with 2D/G relative intensities clearly identify graphene monolayers.[34, 35]

The cantilever beam technique has been employed for subjecting tensile/compressive loads to graphene monolayers (see Fig. 1A). The beam can be flexed up or down by means of an adjustable screw subjecting the flake to compressive or tensile loads, respectively. The maximum deflection of the neutral axis of the beam (elastic behaviour), is given by the following equation (for more details see [16]):

$$\varepsilon(x) = \frac{3t\delta}{2L^2}\left(1 - \frac{x}{L}\right) \qquad (5)$$

where $L$ is the cantilever beam span, $\delta$ is the deflection of the beam (at the free end) at each increment of flexure and $t$ is the beam thickness. The position where Raman measurements are taken is denoted by the variable "$x$". For the above equation to be valid, the span to maximum deflection aspect ratio should be greater than 10.[28]

It has to be noted, it is extremely important to apply the stress smoothly in order to ensure reproducibility and no slippage.[13] In our typical experiments, the strain increment was 0.03 or 0.05%. The maximum strain achieved was usually close to 0.7% due to limitations originating mainly from the mechanical response of the substrate. At this strain level, cracks appeared in both underlying SU8 and overlying S1805. Therefore, to minimize the risk of influencing the results by an imperfect stress transfer to graphene or even the danger of irrecoverably damaging the specimen too early, the experiments were stopped at this point. Nevertheless, as shown, the most important features of the behavior of graphene under small strains (<1.5%) can be deduced from its evolution in the range of our experiments. The data presented in this study were measured on two different flakes (on different beams) on several points in each flake (Fig. 1), sometimes in repeated tension and compression cycles. Altogether, more than 100 and 50 spectra were acquired under compression and tension, respectively.

ACKNOWLEDGMENT

FORTH / ICE-HT acknowledge financial support from the Marie-Curie Transfer of Knowledge program CNTCOMP [Contract No.: MTKD-CT-2005-029876]. K.S.N. is grateful to the Royal Society and European Research Council (grant 207354 – "Graphene") for support.

REFERENCES

(1)     Geim, A. K.; Novoselov, K. S. The rise of graphene. *Nat. Mater.* **2007,** *6*, 183-191.

(2)     Novoselov, K. S.; Geim, A. K.; Morozov, S. V.; Jiang, D.; Zhang, Y.; Dubonos, S. V.; Grigorieva, I. V.; Firsov, A. A. Electric field effect in atomically thin carbon films. *Science* **2004,** *306*, 666-669.

(3)     Lee, C.; Wei, X. D.; Kysar, J. W.; Hone, J. Measurement of the elastic properties and intrinsic strength of monolayer graphene. *Science* **2008,** *321*, 385-388.

(4)     Zhao, Q.; Nardelli, M. B.; Bernholc, J. Ultimate strength of carbon nanotubes: A theoretical study. *Phys. Rev. B* **2002,** *65*, 144105.

(5)     Neto, A. H. C.; Guinea, F.; Peres, N. M. R.; Novoselov, K. S.; Geim, A. K. The electronic properties of graphene. *Rev. Mod. Phys.* **2009,** *81*, 109-54.

(6)     Guinea, F.; Horovitz, B.; Le Doussal, P. Gauge fields, ripples and wrinkles in graphene layers. *Solid State Commun.* **2009,** *149*, 1140-1143.

(7)     Pereira, V. M.; Neto, A. H. C. Strain Engineering of Graphene's Electronic Structure. *Phys. Rev. Lett.* **2009,** *103*, 046801-4.

(8)     Ramanathan, T.; Abdala, A. A.; Stankovich, S.; Dikin, D. A.; Herrera-Alonso, M.; Piner, R. D.; Adamson, D. H.; Schniepp, H. C.; Chen, X.; Ruoff, R. S., et al. Functionalized graphene sheets for polymer nanocomposites. *Nat. Nanotechnol.* **2008,** *3*, 327-331.

(9)     Stankovich, S.; Dikin, D. A.; Dommett, G. H. B.; Kohlhaas, K. M.; Zimney, E. J.; Stach, E. A.; Piner, R. D.; Nguyen, S. T.; Ruoff, R. S. Graphene-based composite materials. *Nature* **2006,** *442*, 282-286.

(10)    Rafiee, M. A.; Rafiee, J.; Yu, Z. Z.; Koratkar, N. Buckling resistant graphene nanocomposites. *Appl. Phys. Lett.* **2009,** *95*, 223103-3.

(11)    Schadler, L. S.; Galiotis, C. Fundamentals and applications of micro-Raman spectroscopy to strain measurements in fibre-reinforced composites. *Int. Mater. Rev.* **1995,** *40*, 116-134.

(12)    Huang, M. Y.; Yan, H. G.; Chen, C. Y.; Song, D. H.; Heinz, T. F.; Hone, J. Phonon softening and crystallographic orientation of strained graphene studied by Raman spectroscopy. *Proc. Natl. Acad. Sci. U.S.A.* **2009,** *106*, 7304-7308.

(13)    Mohiuddin, T. M. G.; Lombardo, A.; Nair, R. R.; Bonetti, A.; Savini, G.; Jalil, R.; Bonini, N.; Basko, D. M.; Galiotis, C.; Marzari, N., et al. Uniaxial strain in graphene by Raman spectroscopy: G peak splitting, Grueneisen parameters, and sample orientation. *Phys. Rev. B* **2009,** *79*, 205433-8.

(14)    Ni, Z. H.; Yu, T.; Lu, Y. H.; Wang, Y. Y.; Feng, Y. P.; Shen, Z. X. Uniaxial Strain on Graphene: Raman Spectroscopy Study and Band-Gap Opening. *Acs Nano* **2008,** *2*, 2301-2305.

(15)    Yu, T.; Ni, Z. H.; Du, C. L.; You, Y. M.; Wang, Y. Y.; Shen, Z. X. Raman mapping investigation of graphene on transparent flexible substrate: The strain effect. *J. Phys. Chem. C* **2008,** *112*, 12602-12605.

(16)    Tsoukleri, G.; Parthenios, J.; Papagelis, K.; Jalil, R.; Ferrari, A. C.; Geim, A. K.; Novoselov, K. S.; Galiotis, C. Subjecting a Graphene Monolayer to Tension and Compression. *Small* **2009,** *5*, 2397-2402.

(17)    Hanfland, M.; Beister, H.; Syassen, K. Graphite under pressure: Equation of state and first-order Raman modes. *Phys. Rev. B* **1989,** *39*, 12598.

(18)    Proctor, J. E.; Gregoryanz, E.; Novoselov, K. S.; Lotya, M.; Coleman, J. N.; Halsall, M. P. High-pressure Raman spectroscopy of graphene. *Phys. Rev. B* **2009,** *80*, 073408-4.

(19)    Bunch, J. S.; van der Zande, A. M.; Verbridge, S. S.; Frank, I. W.; Tanenbaum, D. M.; Parpia, J. M.; Craighead, H. G.; McEuen, P. L. Electromechanical Resonators from Graphene Sheets. *Science* **2007,** *315*, 490-493.

(20)    Stolyarova, E.; Stolyarov, D.; Bolotin, K.; Ryu, S.; Liu, L.; Rim, K. T.; Klima, M.; Hybertsen, M.; Pogorelsky, I.; Pavlishin, I., et al. Observation of Graphene Bubbles and Effective Mass Transport under Graphene Films. *Nano Lett.* **2008,** *9*, 332-337.

(21)    Metzger, C.; Remi, S.; Liu, M.; Kusminskiy, S. V.; Castro Neto, A. H.; Swan, A. K.; Goldberg, B. B. Biaxial Strain in Graphene Adhered to Shallow Depressions. *Nano Lett.* **2009,** *10*, 6-10.

(22)   Li, Z.; Cheng, Z.; Wang, R.; Li, Q.; Fang, Y. Spontaneous Formation of Nanostructures in Graphene. *Nano Lett.* **2009,** *9*, 3599-3602.

(23)   Atalaya, J.; Isacsson, A.; Kinaret, J. M. Continuum Elastic Modeling of Graphene Resonators. *Nano Lett.* **2008,** *8*, 4196-4200.

(24)   Gao, Y.; Hao, P. Mechanical properties of monolayer graphene under tensile and compressive loading. *Phys. E* **2009,** *41*, 1561-1566.

(25)   Scarpa, F.; Adhikari, S.; Phani, A. S. Effective elastic mechanical properties of single layer graphene sheets. *Nanotechnology* **2009,** *20*, 065709.

(26)   Ferralis, N.; Maboudian, R.; Carraro, C. Evidence of Structural Strain in Epitaxial Graphene Layers on 6H-SiC(0001). *Phys. Rev. Lett.* **2008,** *101*.

(27)   Robinson, J. A.; Puls, C. P.; Staley, N. E.; Stitt, J. P.; Fanton, M. A.; Emtsev, K. V.; Seyller, T.; Liu, Y. Raman Topography and Strain Uniformity of Large-Area Epitaxial Graphene. *Nano Lett.* **2009,** *9*, 964-968.

(28)   Timoshenko, S. P.; Gere, J. M., *Theory of Elastic Stability*. McGraw-Hill: New York, 1961.

(29)   Lanir, Y.; Fung, Y. C. B. Fiber Composite Columns under Compression. *Journal of Composite Materials* **1972,** *6*, 387-&.

(30)   Fasolino, A.; Los, J. H.; Katsnelson, M. I. Intrinsic ripples in graphene. *Nat. Mater.* **2007,** *6*, 858-861.

(31)   Lourie, O.; Cox, D. M.; Wagner, H. D. Buckling and collapse of embedded carbon nanotubes. *Phys. Rev. Lett.* **1998,** *81*, 1638-1641.

(32)   Chen, C.-C.; Bao, W.; Theiss, J.; Dames, C.; Lau, C. N.; Cronin, S. B. Raman Spectroscopy of Ripple Formation in Suspended Graphene. *Nano Lett.* **2009,** *9*, 4172-4176.

(33)   Melanitis, N.; Tetlow, P. L.; Galiotis, C.; Smith, S. B. Compressional Behavior of Carbon-Fibers .2. Modulus Softening. *J. Mater. Sci.* **1994,** *29*, 786-799.

(34)	Ferrari, A. C.; Meyer, J. C.; Scardaci, V.; Casiraghi, C.; Lazzeri, M.; Mauri, F.; Piscanec, S.; Jiang, D.; Novoselov, K. S.; Roth, S., et al. Raman spectrum of graphene and graphene layers. *Phys. Rev. Lett.* **2006,** *97*, 187401.

(35)	Malard, L. M.; Pimenta, M. A.; Dresselhaus, G.; Dresselhaus, M. S. Raman spectroscopy in graphene. *Phys. Rep.* **2009,** *473*, 51-87.

# Compression Behavior of Single-layer Graphene

*Otakar Frank[1], Georgia Tsoukleri[1,2], John Parthenios[1,2], Konstantinos Papagelis[3], Ibtsam Riaz[4], Rashid Jalil[4], Kostya S. Novoselov[4], and Costas Galiotis[1,2,3]*

## Supporting Information

**Raman study of graphene - background**

The recently developed method for graphene preparation by micromechanical cleavage of graphite[1] provides an opportunity for studying the Raman band shifts of both G and 2D modes[2,3] upon tensile or compressive loading at the molecular level.[4-9] This is important not only for highlighting the extreme strength and stiffness of graphene but also to link its behaviour with the mechanical deformation of other graphitic structures such as bulk graphite, carbon nanotubes (CNT) and CF. The G peak corresponds to the doubly degenerate $E_{2g}$ phonon at the Brillouin zone centre. The D peak is due to the breathing modes of $sp^2$ rings and requires a defect for its activation.[2,3,10] It comes from TO phonons around the **K** point of the Brillouin zone, is active by double resonance[11] and is strongly dispersive with excitation energy due to a Kohn Anomaly at **K**.[12] The 2D peak is the second order of the D peak. This is a single peak in monolayer graphene, whereas it splits in four in bilayer graphene, reflecting the evolution of the band structure.[2,3] Since the 2D peak originates from a process where momentum conservation is obtained by the participation of two phonons with opposite wavevectors it does not require the presence of defects for its activation, and is thus always present. Indeed, high quality graphene shows the G, 2D peaks, but not D.

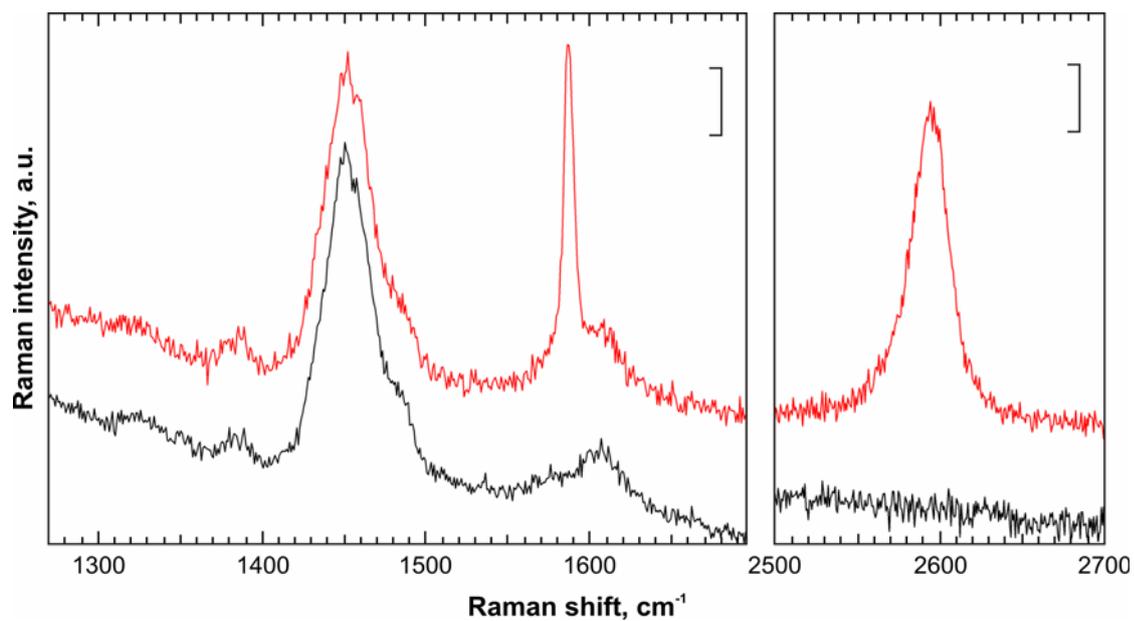

**Figure S1.** Original Raman spectra excited by a 785 nm laser of the combined SU8 and S1805 substrate (black) and a graphene flake embedded within this substrate (red). The spectra are offset for clarity, and scale bars represent 500 counts in both panels.

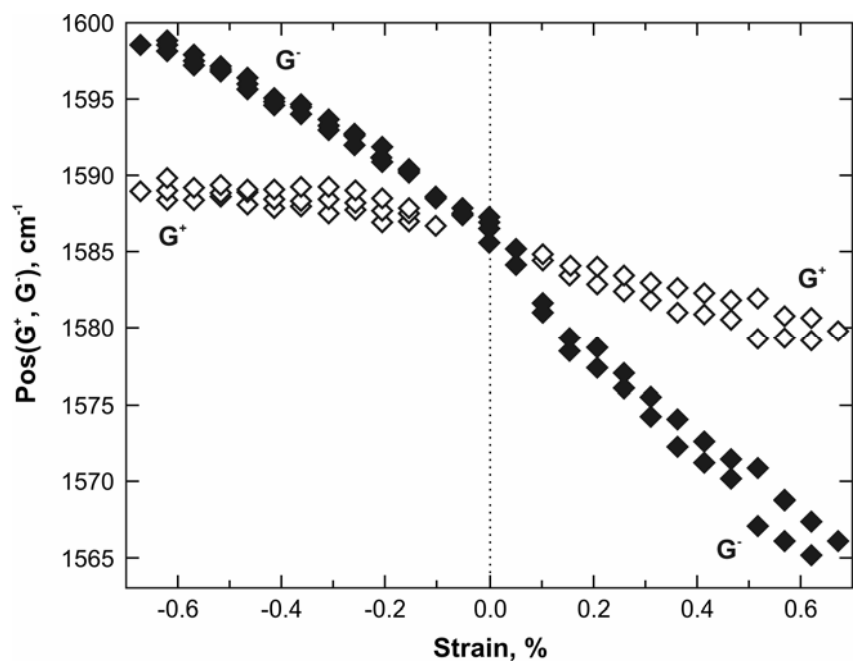

**Figure S2.** Plot of G band positions as a function of strain from experiments conducted of flake F1 (Figure 1c, main text). Strain with positive (negative) values indicates tension (compression). Full end empty diamonds indicate the frequency position of the G⁻ and G⁺ sub-bands.

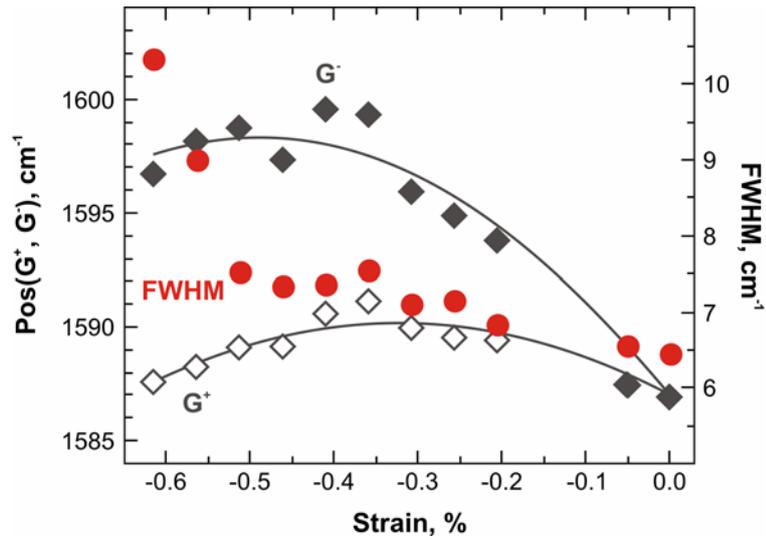

**Figure S3.** Plot of G⁻ and G⁺ band positions and their FWHM as a function of strain on a selected spot on flake F2 (Fig. 1c, main text). Full circles indicate the bands' FWHM (right axis), full (empty) diamonds show the position of the G⁻ (G⁺) sub-bands. Only one set of FWHM is presented, since both sub-bands have the same width for a given strain level. Solid (dashed) lines are 2$^{nd}$ order polynomial fits of the G⁻(G⁺) band position measurements.

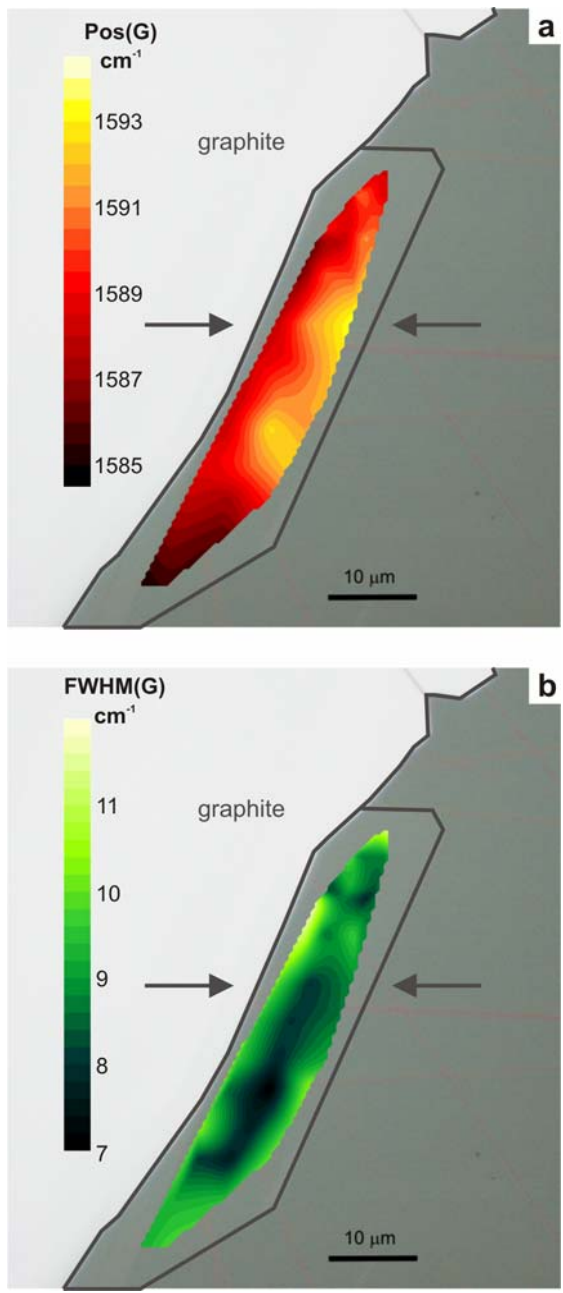

**Figure S4.** Post mortem (a) Pos(G) and (b) FWHM(G) maps of specimen F2 after cyclic loading. The band was fitted as a single Lorentzian. The light grey area in both (a) and (b) corresponds to bulk graphite. The arrows indicate the strain direction.

|  | G⁻ | | | G⁺ | |
|---|---|---|---|---|---|
|  | $a_0$ | $a_1$ | $a_2$ | $a_1$ | $a_2$ |
| **F1** | | | | | |
| 1 | 1586.9 | -22.5 ± 0.9 | -5.9 ± 1.8 | -1.8 ± 1.0 | 1.4 ± 1.9 |
| 2 | 1587.3 | -18.8 ± 0.8 | -2.5 ± 1.6 | -4.0 ± 0.9 | -1.94 ± 1.7 |
| 3 | 1586.5 | -25.2 ± 1.0 | -9.9 ± 2.0 | -10.5 ± 1.2 | -9.54 ± 2.4 |
| **F2** | | | | | |
| 1 | 1585.1 | -36.1 ± 1.8 | -28.9 ± 3.8 | -12.2 ± 0.8 | -17.14 ± 1.6 |
| 2 | 1584.6 | -34.1 ± 2.8 | -24.1 ± 6.1 | -13.0 ± 1.4 | -18.2 ± 3.1 |
| 3 | 1583.3 | -28.4 ± 1.0 | -13.0 ± 2.2 | -4.5 ± 1.0 | -0.7 ± 2.2 |

|  | 2D | | |
|---|---|---|---|
|  | $a_0$ | $a_1$ | $a_2$ |
| **F1** | | | |
| 1 | 2956.5 | -38.0 ± 2.1 | -10.3 ± 4.1 |
| 2 | 2596.6 | -36.6 ± 3.4 | -13.5 ± 6.3 |
| 3 | 2594.6 | -41.7 ± 3.2 | -12.6 ± 6.4 |
| **F2** | | | |
| 1 | 2592.3 | -59.8 ± 4.5 | -52.4 ± 9.8 |
| 2 | 2591.3 | -60.1 ± 7.6 | -54.8 ± 16.6 |
| 3 | 2598.4 | -45.4 ± 7.8 | -19.9 ± 17.2 |

**Table S1.** Coefficients of 2$^{nd}$ order polynomial curves fitted to the Raman G$^+$, G$^-$ and 2D bands evolution with compressive strain. The fit equation can be written: $\omega = a_0 + a_1 |\varepsilon| - a_2 \varepsilon^2$. Thus $a_0 = \omega^0$ (cm$^{-1}$), $a_1$ corresponds to the strain sensitivity $\partial_{G,2D}/\partial \varepsilon$ (cm$^{-1}$/%) close to zero strain, and $a_2$ expresses the curvature of the slope $\partial_{G,2D}/\partial \varepsilon^2$ (cm$^{-1}$/%$^2$). The ± sign prefaces a value of 95% confidence interval.

**Supporting Information References**

1. Novoselov, K. S.; Geim, A. K.; Morozov, S. V.; Jiang, D.; Zhang, Y.; Dubonos, S. V.; Grigorieva, I. V.; Firsov, A. A. *Science* **2004,** *306* (5296), 666-669.

2. Ferrari, A. C.; Meyer, J. C.; Scardaci, V.; Casiraghi, C.; Lazzeri, M.; Mauri, F.; Piscanec, S.; Jiang, D.; Novoselov, K. S.; Roth, S.; Geim, A. K. *Phys. Rev. Lett.* **2006,** *97* (18), 187401.

3. Malard, L. M.; Pimenta, M. A.; Dresselhaus, G.; Dresselhaus, M. S. *Phys. Rep.* **2009,** *473* (5-6), 51-87.

4. Huang, M. Y.; Yan, H. G.; Chen, C. Y.; Song, D. H.; Heinz, T. F.; Hone, J. *Proc. Natl. Acad. Sci. U.S.A.* **2009,** *106* (18), 7304-7308.

5. Mohiuddin, T. M. G.; Lombardo, A.; Nair, R. R.; Bonetti, A.; Savini, G.; Jalil, R.; Bonini, N.; Basko, D. M.; Galiotis, C.; Marzari, N.; Novoselov, K. S.; Geim, A. K.; Ferrari, A. C. *Phys. Rev. B* **2009,** *79* (20), 205433-8.

6. Ni, Z. H.; Yu, T.; Lu, Y. H.; Wang, Y. Y.; Feng, Y. P.; Shen, Z. X. *Acs Nano* **2008,** *2* (11), 2301-2305.

7. Proctor, J. E.; Gregoryanz, E.; Novoselov, K. S.; Lotya, M.; Coleman, J. N.; Halsall, M. P. *Phys. Rev. B* **2009,** *80* (7), 073408-4.

8. Yu, T.; Ni, Z. H.; Du, C. L.; You, Y. M.; Wang, Y. Y.; Shen, Z. X. *J. Phys. Chem. C* **2008,** *112* (33), 12602-12605.

9. Tsoukleri, G.; Parthenios, J.; Papagelis, K.; Jalil, R.; Ferrari, A. C.; Geim, A. K.; Novoselov, K. S.; Galiotis, C. *Small* **2009,** *5* (21), 2397-2402.

10. Tuinstra, F.; Koenig, J. L. *The Journal of Chemical Physics* **1970,** *53* (3), 1126-1130.

11. Maultzsch, J.; Reich, S.; Thomsen, C. *Phys. Rev. B* **2004,** *70* (15), 155403.

12. Piscanec, S.; Lazzeri, M.; Mauri, F.; Ferrari, A. C.; Robertson, J. *Phys. Rev. Lett.* **2004,** *93* (18).